\documentclass[12pt]{article}
\usepackage{epsf} 

\begin{document}

\baselineskip 18pt

\newcommand{\sheptitle}
{Hagedorn Inflation: 
Open Strings on Branes Can Drive Inflation}

\newcommand{\shepauthor}
{
Steven Abel$^{a,}$\footnote[1]{S.A.Abel@durham.ac.uk},
Katherine Freese $^{b,}$\footnote[2]{ktfreese@umich.edu}, 
Ian I.~Kogan $^{c,}$\footnote[3]{i.kogan@physics.ox.ac.uk}}

\newcommand{\shepaddress}
{
$a$ Dept. of Mathematical Sciences, University of Durham, 
	Science Laboratories, \hfil\break
	South Rd., Durham DH1 3LE, United Kingdom\\
$b$ Michigan Center for Theoretical Physics,
        University of Michigan, Ann Arbor MI 48109-1120, USA\\
$c$ Theoretical Physics, 1 Keble Rd, Oxford OX1 3NP, UK}

\newcommand{\shepabstract}
{We demonstrate an inflationary solution to the cosmological horizon
problem during the Hagedorn regime in the
early universe.  Here the observable universe is confined to three
spatial dimensions (a three-brane) embedded in higher dimensions.
The only ingredients required are open strings on
D-branes at temperatures close to the string scale.
No potential is required.
Winding modes of the strings
provide a negative pressure that can drive inflation of our
observable universe.  Hence the mere existence of open strings on branes
in the early hot phase of the universe drives Hagedorn inflation,
which can be either power law or exponential. 
We note the amusing fact that, in the case of  stationary
extra dimensions, inflationary expansion takes place only for branes
of three or less dimensions.}

\begin{titlepage}
\begin{flushright}
MCTP-03-10

\end{flushright}
\vspace{0.5in}
\begin{center}
{\large{\bf \sheptitle}}
\bigskip \\ \shepauthor \\ \mbox{} \\ {\it \shepaddress} \\ 
\vspace{0.5in}
{\bf Abstract} \bigskip \end{center} \setcounter{page}{0}
\shepabstract
\end{titlepage}

\def\sspace{\baselineskip = .16in}
\def\dspace{\baselineskip = .30in}
\def\beq{\begin{equation}}
\def\eeq{\end{equation}}
\def\bea{\begin{eqnarray}}
\def\eea{\end{eqnarray}}
\def\bq{\begin{quote}}
\def\eq{\end{quote}}
\def\ra{\rightarrow}
\def\lra{\leftrightarrow}
\def\ups{\upsilon}
\def\bq{\begin{quote}}
\def\eq{\end{quote}}
\def\ra{\rightarrow}
\def\un{\underline}
\def\ov{\overline}
\def\ord{\cal O} 

\newcommand{\plb}[3]{{{\it Phys.~Lett.}~{\bf B#1} (#3) #2}}
\newcommand{\npb}[3]{{{\it Nucl.~Phys.}~{\bf B#1} (#3) #2}}
\newcommand{\prd}[3]{{{\it Phys.~Rev.}~{\bf D#1} (#3) #2}}
\newcommand{\ptp}[3]{{{\it Prog.~Theor.~Phys.}~{\bf #1} (#3) #2}}
\newcommand{\ijmpa}[3]{{{\it Int.~J.~Mod.~Phys.}~{\bf A#1} (#3) #2}}
\newcommand{\prl}[3]{{{\it Phys.~Rev.~Lett.}~{\bf #1} (#3) #2}}
\newcommand{\hepph}[1]{{\tt hep-ph/#1}}
\newcommand{\hepth}[1]{{\tt hep-th/#1}}
\newcommand{\grqc}[1]{{\tt gr-qc/#1}} 
\newcommand{\leqsim}{\,\raisebox{-0.6ex}{$\buildrel < \over \sim$}\,}
\newcommand{\geqsim}{\,\raisebox{-0.6ex}{$\buildrel > \over \sim$}\,}
\newcommand{\nin}{\,  \mbox{$/$ \hspace{-0.2cm} $\in$}\,}
\newcommand{\be}{\begin{equation}}
\newcommand{\ee}{\end{equation}}
\newcommand{\ba}{\begin{eqnarray}}
\newcommand{\ea}{\end{eqnarray}}
\newcommand{\nn}{\nonumber}
\newcommand{\cf}{\mbox{{\em c.f.~}}}
\newcommand{\ie}{\mbox{{\em i.e.~}}}
\newcommand{\eg}{\mbox{{\em e.g.~}}}
\newcommand{\mpl}{\mbox{$M_{pl}$}}
\newcommand{\ol}[1]{\overline{#1}}
\newcommand{\eqr}[1]{eq.(\ref{#1})}
\def\gev{\,{\rm GeV }}
\def\tev{\,{\rm TeV }}
\def\dd{\mbox{d}}
\def\etal{\mbox{\it et al }}
\def\half{\frac{1}{2}}
\def\Tr{\mbox{Tr}}
\def\bra{\langle}
\def\ket{\rangle}
\def\lim{\mbox{{\bf L}} }
\def\nlim{\mbox{{\bf NL}} }
\def\sclim{\mbox{\tiny{\bf L}} }
\def\scnlim{\mbox{\tiny{\bf NL}} }
\def\nlimc{\mbox{{\bf NL$_{closed}$}} }
\def\nlimo{\mbox{{\bf NL$_{open}$}} }
\def\Vp{V_{\parallel}}
\def\Vt{V_{\perp}} 
\def\vp{V_{\parallel}}
\def\vt{V_{\perp}} 
\def\wp{W_{\parallel}}
\def\wt{W_{\perp}} 
\def\Wp{W_{\parallel}}
\def\Wt{W_{\perp}} 
\def\Rt{R_{\perp}}
\def\ep{\varepsilon} 
\newcommand{\smallfrac}[2]{\frac{\mbox{\small #1}}{\mbox{\small #2}}}

\def\CAG{{\cal A/\cal G}}           \def\CO{{\cal O}} \def\CZ{{\cal Z}}
\def\CA{{\cal A}} \def\CC{{\cal C}} \def\CF{{\cal F}} \def\CG{{\cal G}}
\def\CL{{\cal L}} \def\CH{{\cal H}} \def\CI{{\cal I}} \def\CU{{\cal U}}
\def\CB{{\cal B}} \def\CR{{\cal R}} \def\CD{{\cal D}} \def\CT{{\cal T}}
\def\CM{{\cal M}} \def\CP{{\cal P}}
\def\CN{{\cal N}} \def\CS{{\cal S}}  

\section{Introduction}
\label{sec:intro}

	Inflationary cosmology \cite{guth} was proposed as a solution to the
horizon, flatness, and monopole problems of the standard Hot Big Bang
scenario. Inflation requires a period of accelerated expansion,
$\ddot a >0$, corresponding to a superluminal expansion of the
scale factor.  In the standard 3+1 dimensional universe, the 
Friedmann Robertson Walker (FRW) equations imply
${\ddot a \over a} = - {4 \pi \over 3 m_{pl}^2} (\rho + 3p)$.
Hence accelerated expansion is provided by a negative pressure.
In standard inflationary models, this negative pressure is
provided by a vacuum energy (a potential) with $p = - \rho$.

We have found \cite{us} an entirely different source of
negative pressure:  open strings on D-branes
at temperatures close to the string scale. 
Although Einstein's equations in higher dimensions take a different
form than the FRW equation above, negative pressure can still
drive inflation. Note that there is no potential of any kind
in our model; instead, open strings on branes drive the inflation.

At sufficiently high temperatures and densities 
fundamental strings enter a curious `long string' 
Hagedorn phase~\cite{carlitz,general,deo,thorl,abkr,after}.  A
classical random walk picture can be used
to model the behaviour of the strings in 
cosmological backgrounds. The particular systems we will focus on 
are D-branes in the weak coupling limit~\cite{polch}. 
In particular, we consider the scenario in which our observable
universe is confined to three spatial dimensions (a three-brane)
embedded in higher dimensions.  We will denote the
rest of the universe outside of our 3-brane as the bulk.  We can
separate the energy momentum tensor into two components:  a 
localized component corresponding to the D-brane tension, and 
a diffuse component that spreads into the bulk corresponding to 
open string excitations of the brane. 
We find an interesting type of cosmological effect of a
primordial Hagedorn phase of open strings on branes: 

\begin{itemize}

\item {\it Hagedorn inflation}.  The transverse 
`bulk' components of the energy-momentum tensor can be negative. 
If all of the transverse dimensions have winding modes,
this negative `pressure' causes the brane to 
power law inflate along its length with 
a scale factor that varies as $a\sim t^{4/3}$
even in the absence of a nett cosmological constant 
(see eq.(\ref{61})).
If there are transverse dimensions that are large (in the sense that 
the string modes are not space-filling in these directions),
then we can find exponential inflation (see eq.(\ref{expinf})). 
{\it No potential is required here}. Merely the existence of open strings
on D-branes drives the inflation.

\end{itemize}

We begin by discussing string thermodynamics of open strings on
D-branes at high temperatures near the string scale.  We derive the
energy-momentum tensor required for Einstein's equations from the
partition function, which in turn can be derived from the density of
states in a random walk approach.  Winding modes of the strings give
rise to a negative `pressure' in the bulk (the directions
perpendicular to the three-brane on which we live).  Armed with the
energy-momentum tensor in Eq.(\ref{eq:imp}), we examine the resultant
cosmology. We can solve Einstein's equations with various ans\"atze in
the presence of this negative bulk component. Our primary result is
that we find Hagedorn inflation of our observable universe due to the
negative pressure in the bulk.

If one assumes  adiabaticity, the inflationary growth period 
drives down the temperature of the system; eventually
the temperature drop causes
the universe to leave the Hagedorn regime, and consequently
inflation ends automatically.  In fact our solutions are only
valid for small changes in the metric corresponding to small
changes in the volumes, i.e., we can demonstrate instability
to inflationary expansion but cannot follow the solutions further.
If one uses the solutions beyond
the region of their validity, the period of superluminal
growth ends too quickly to solve cosmological problems.
Hence we do also discuss how, in non-adiabatic systems,
inflation can be sustained.

An amusing result is that power law
inflation in the universal high energy system only 
takes place if the number of large dimensions on the brane
is $p\leq 3$. In addition, branes tend to `melt' unless
$p \leq 4$. Hence one can speculate on the role of these
effects in the fact that our observable universe has three
large dimensions.

\section{The Hagedorn phase and random walks} 
 
We here derive the thermodynamic properties of D-branes
using a random walk approach.  The results
are independent of the details of the
compactification and the degree of supersymmetry,
and agree with previous work \cite{abkr} assuming toroidal
compactification in microcanonical and canonical ensembles.
The reader interested only in the result for the energy-momentum
tensor and the resultant cosmology should proceed directly
to Eq.(\ref{eq:imp}).

\subsection{String thermodynamics and the Hagedorn phase}

The Hagedorn phase arises in theories containing fundamental
strings because they have a large number of internal degrees of 
freedom. Because of the existence of many 
oscillator modes, the density of states 
grows exponentially with single string energy $\varepsilon$, $\omega(\varepsilon) 
\sim \varepsilon^{-b}
e^{\beta_H \varepsilon}$, where the inverse Hagedorn temperature $\beta_H$ 
(where $\beta =1/T$)
and the exponent $b$ depend on the particular
theory in question (for example heterotic or type II)~\cite{carlitz}. 
For type I,IIA,IIB strings the numerical value of the 
inverse Hagedorn temperature
is $\beta_H=2 \sqrt{2} \pi$ in string units.
Thermodynamic quantities, such as the 
entropy, diverge at the 
Hagedorn temperature.
The partition function $Z$ is obtained 
as the integral of the density of states times the
 usual Boltzmann factor $e^{-\beta \varepsilon}$,
$Z \propto \int d\varepsilon \, \varepsilon^{-b} \, e^{(\beta_H - \beta) 
\varepsilon}$ which diverges at
$\beta = \beta_H$ for $b \leq 2$. 

In the Hagedorn regime, fundamental strings can be described
as `long strings' in a random walk analogy. 
The size of the random walk (the distance covered by the string)
is given by the length scale $\sqrt{\varepsilon}$. The physical setup
is shown in Figure 1.  

\begin{figure}[htp]
\begin{center}
\hspace*{0.5in}
\epsfxsize=5.0in
\epsffile{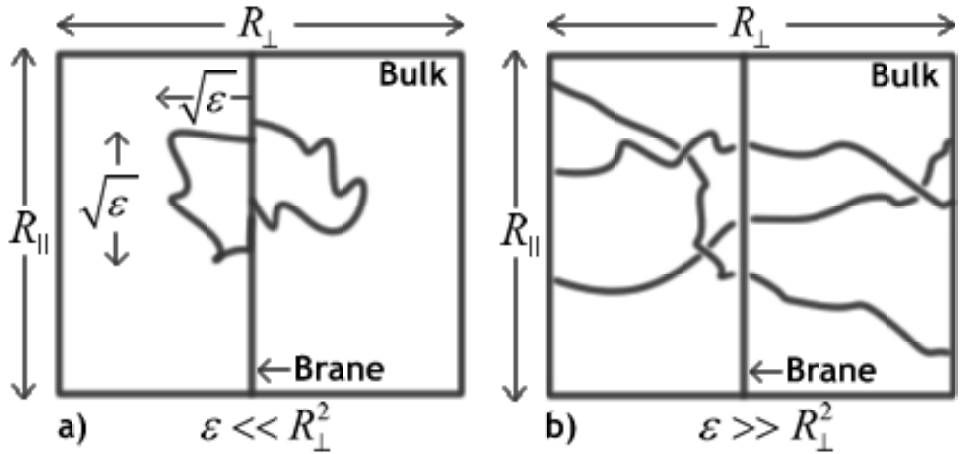}
\caption{We consider a volume, portrayed by the box,
containing a single brane embedded in the bulk.
The quantity $R_{||}$
indicates the size of the dimensions parallel to the brane 
(for cosmology, $R_{||} \rightarrow \infty$) and $R_{\perp}$
indicates the size of the dimensions perpendicular to the brane.
The random walk of strings attached to the brane 
traverses a distance $\sqrt{\varepsilon}$.
Figs. 1a represents the case $\sqrt{\epsilon} << R_{\perp}$.
Fig. 1b represents the case  
$\sqrt{\epsilon} >> R_{\perp}$.  This latter case is our standard
high energy case in which all modes are space-filling
(effectively, ``winding modes'').   }
\end{center}
\end{figure}

We consider a volume, portrayed by the box,
containing a single brane.  The quantity $R_{||}$
indicates the size of the dimensions parallel to the brane 
(for cosmology, $R_{||} \rightarrow \infty$) and $R_{\perp}$
indicates the size of the dimensions perpendicular to the brane.
The random walk of strings attached to the brane 
traverses a distance $\sqrt{\varepsilon}$.
Figs. 1a and 1b show the cases $\sqrt{\epsilon} << R_{\perp}$ and 
$\sqrt{\epsilon} >> R_{\perp}$ respectively.  
In this paper we are particularly
interested in the high energy case of fig. 1b, in which the strings
fill all the space; for the case of toroidal compactification
this corresponds to winding modes.  We will use the nomenclature
`winding modes' to encompass these space-filling modes regardless
of the type of compactification.  We define $d_o$ to be the number
of dimensions transverse to the brane in which there are {\it no windings}.
As our standard `high-energy' regime, we will take the case of $d_o = 0$, 
so that all dimensions have windings;
this is the system which is always reached provided that the 
the energy density is high enough.

Previously, \cite{abkr}  obtained
 the distribution function $\omega (\ep)$ for
open strings attached to 
a brane:
\be
\omega (\ep)_{\rm open} \sim {\Vp \over  
\ep^{d_\perp /2}} \;\exp\,(\beta_H\,\ep) \,\,\hspace{1cm} 
R_{\perp} \gg \sqrt{\ep} \, .
\ee 
and
\be
\label{eq:flatspace}
\omega(\ep)_{\rm open} \sim {\Vp \over \Vt}  
\;\exp\,(\beta_H \,\ep) \,\,\hspace{1cm} 
R_{\perp} \ll \sqrt{\ep} .
\ee 
Here, $d_\perp$ is the number of dimensions perpendicular to the brane,
and $\Vp$ and $\Vt$ are the volumes parallel and perpendicular to the brane
respectively.

One can now adapt the random walk to a cosmological background.
In the case of a non-trivial 
metric the most natural interpretation of the parameter $\ep$ is that it is 
the proper length of 
a string in the bulk and certainly we can always go 
to the local inertial frame in which a small portion of the string 
has the usual Euclidean energy~$\equiv$~length equivalence. 

We make the usual quasi-equilibrium approximation that equilibrium is
established much more quickly than any change in the metric so that
the metric may be taken to be approximately constant when evaluating
properties such as density.  We assume that the metric is expressed in
terms of parallel dimensions $x$ and transverse ones $y$ \be ds^2 =
-n^2 dt^2 + g_{||ij}dx^idx^j +g_{\perp nm}dy^ndy^m , \ee with the
brane lying at $y^n=0$.  We      define an averaging over the extra
dimension with an overbar, \be
\label{eq:avgquant}
\overline{O} = \frac{\int dy \sqrt{g_\perp} O(y) }{\int dy'
\sqrt{g_\perp } }\, .
\ee
The density of states for the limiting systems is found
to be the same expression as in the flat space case of eq.
(\ref{eq:flatspace})
but with all volumes averaged over transverse dimensions as in 
eq.(\ref{eq:avgquant}).

\subsection{Partition Function}

Now that we have obtained a density of states of open strings attached
to branes near the Hagedorn temperature, we can find
the partition function at temperature $1/\beta$,
\be
\log Z(\beta, \overline{\vp},\overline{\vt}) \sim \int d\varepsilon
\,\, \omega(\varepsilon) \,\, e^{-\beta \varepsilon} .
\ee
In the standard $d_o=0$ case,
\be
\label{eq:partition}
\log Z(\beta,\overline{\Vp},\overline{\Vt} ) = 
2 \frac{\overline{\Vp}^2 \beta_H^2}{\overline{\Vt} (\beta^2-\beta_H^2) } 
+ {\rm nonsingular} \,\, {\rm cutoff}\,\, {\rm terms} 
\ee
which are the successive terms in a saddle point approximation.
In \cite{us}, we present the density of states and partition
function for arbitrary values of $d_o$ as well.

\section{Stress-energy tensor $T_{\mu\nu}$ in a bulk Hagedorn phase}

We now use these thermodynamic results to 
find the bulk energy momentum tensor during the Hagedorn regime.
We may find the energy momentum tensor from 
\be 
\label{eq:tuv}
\langle T^\mu_\nu \rangle = 2 \frac{g^{\mu\rho}}{\sqrt{g}} 
\frac{\delta \log Z(\beta,\overline{\Vp},\overline{\Vt}) } 
{\delta g^{\rho\nu}}.
\ee  
For now, we take a single extra dimensions,
so that $\mu,\nu = 0,1,2,3,5$ and $5$ is the extra dimension.
We will treat the functional derivative with respect to $g_{\mu\nu}$
in the following way.  We assume that small changes in the metric 
correspond to making small changes in the volumes in $\log Z$, 
\eg for a single extra dimension
\be
\label{eq:fnal}
{\delta Z \over \delta g_{55}} = \int dx' {\delta Z \over \delta
\overline{V}_\perp(x')} 
{\delta \overline{V}_\perp (x') \over \delta g_{55}} \, .
\ee
Then, in the case of only one extra dimension, we can write 
\be
\label{eq:fnal2}
\overline{\Vt} = \int dy \sqrt{g_{55}} =
{\int d^5x \sqrt{g_{55}} \over \int d^4x}
= {1 \over \Vp \beta} \int d^5x \sqrt{g_{55}} \, ,
\ee
and 
\be 
\label{eq:fnal3}
{\delta \overline{\Vt} \over \delta g_{55}} = \frac{1}{  2
\sqrt{g_{55}}  \Vp \beta} \, .
\ee
Then from eqs.(\ref{eq:fnal}--\ref{eq:fnal3}) we can determine
the functional derivative in eq.(\ref{eq:tuv}).
Our ansatz automatically means that $T_{05} = 0$
and hence $G_{05} = 0$; in other words we are not considering energy exchange
between the brane and the bulk. (In general there might
be energy flux between the two.)

We now summarize the results for the energy-momentum tensor.
We first drop the overline notation of the previous section 
and simply redefine $\Vt$ and $\Vp$ to be the transverse and parallel 
volumes covariantly averaged over the region
of the transverse dimensions covered by the strings.
We define an energy density $\rho$ of strings 
\be 
\rho = \frac{E}{\Vp}
\ee
where $E$ is the total energy of the string system.
We define
\be
\label{eq:defgamma}
\gamma \equiv d_o/2-1
\ee  
where $d_o$ is the number of dimensions with no windings.

We find that the `bulk' components of 
the energy momentum tensor are given by
\ba
\label{eq:imp}
\hat{T}^0_0 & = & -\hat{\rho}
\nonumber \\
\hat{T}^i_i & = & \hat{p}_{\gamma} \nonumber \\
\hat{T}^m_m \equiv p_{bulk} &\approx & \left\{
\begin{array}{ll}
-\hat{p}_{\gamma} & \mbox{transverse with windings}
\\
0 & \mbox{transverse without windings,}
\end{array}\right.
\label{emgamma}
\ea
where $\hat{\rho}=\rho/\Vt$ and 
\be
\label{eq:phat}
\hat{p}_{\gamma} =   
\left\{
\begin{array}{ll}
\frac{1}{\Vt^{3/2}} 
\rho^{\frac{\gamma}{\gamma-1}}
\;\;\;\;\;\;&{\rm if}\;\;\;\;{\gamma=-1,-\half,\half}
  \nn\\
\;\nn\\
\frac{1}{\Vt^{3/2}} 
\log\rho
\;\;\;\;\;\;&{\rm if}\;\;\;\;{\gamma=0}
\nn\\
\,\nn \\
\frac{1}{\Vt^{3/2}} 
e^{-\rho}
\;\;\;\;\;\;&{\rm if}\;\;\;\;{\gamma=1,}
\end{array}
\right.
\ee
wherever there are strings present, and zero otherwise.
In particular, for our standard high energy case of $\gamma = -1$,
the negative bulk pressure is:
\be
\label{eq:pbulk}
p_{bulk} \sim - \rho^{1/2}/\Vt^{3/2} .
\ee

As expected $\hat{T}_0^0$ resembles the local energy density
of strings. The  $\hat{T}^i_i$ represents a relatively small 
pressure coming from Kaluza-Klein modes in the Neumann directions and 
$\hat{T}^m_m$ is a {\em negative} pressure coming from winding modes
in the Dirichlet directions. If we T-dualize the Dirichlet
directions these `winding modes'  
also become Kaluza-Klein modes in Neumann-directions and $T_m^m$ becomes 
positive. Thus negative $T_m^m$ reflects the fact that we have 
T-dualized a dimension much smaller than the string scale thereby 
reversing the pressure. For this reason negative $T_\mu^\mu$ is 
expected to be a 
general feature of space-filling excitations in transverse dimensions. 
The most important result of this section is the negative bulk
pressure found in Eqs.(\ref{emgamma}) and (\ref{eq:pbulk}).

\section{Cosmological Equations in $D=5$}

We now consider
a D-brane configuration that has 3 large parallel 
dimensions (\ie the `observable
universe') and only one transverse dimension $y$ that supports winding 
modes.
For now, we also assume adiabaticity.

We use the metric, 
\be 
\label{simple}
ds^2 = -n^2 dt^2 + a^2 d{\bf x}^2 + b^2 dy^2 ,
\ee
which foliates the space into
flat, homogeneous, and isotropic spatial 3-planes.
Here ${\bf x} = x_1,x_2,x_3$ are the coordinates on the
spatial 3-planes while $y$ is the  coordinate
of the extra dimension.  For simplicity, we make a further
restriction by imposing $Z_2$ symmetry under $y
\rightarrow -y$.  
Without any loss of generality we choose the 3-brane
of the `observable universe' to be fixed at $y=0$.
We can associate a scale 
factor with the parallel dimensions $a(t,y)$ and one for the 
`extra' transverse dimensions $b(t,y)$.
We define $a_0(t(\tau))=a(t,0)$ 
as the scale factor describing the expansion
of the 3-brane where $t(\tau) \equiv \int d\tau n(\tau,y=0)$ is the
proper time of a comoving observer.

Several authors \cite{KKOP} - \cite{chuf}  have
presented the bulk Einstein equations.
For illustrative purposes we restate the results for the 55 (yy) equation:
\be
\hat{G}_{55} = 3\Biggl\{\frac{a'}{a}\,\Biggl(\frac{a'}{a} +
\frac{n'}{n}\Biggr) -\frac{b^2}{n^2}\,\Biggl[\frac{\dot{a}}{a}\,
\Biggl(\frac{\dot{a}}{a}-\frac{\dot{n}}{n}\Biggr) +
\frac{\ddot{a}}{a}\Biggr]\Biggr\} = \hat{\kappa}^2\,\hat{T}_{55}\,, 
\label{55}
\ee
where $\hat{\kappa}^2=8\pi \hat{G}=8\pi/M_5^3$,
$M_5$ is the five-dimensional Planck mass, and the dots and primes
denote differentiation with respect to $t$ and $y$, respectively.
As stated earlier, our ansatz implies that 
$T_{05}=0$ in the bulk.  Shortly we will use the $T_{\mu\nu}$
appropriate to the Hagedorn regime on the right hand side of
Einstein's equations.  

From Eqn.(\ref{55}),
one can see right away
the importance of negative components of the bulk energy momentum tensor.
In the case where time derivatives of $a$ dominate on the left
hand side of the equation, 
one has ${b^2 \over n^2} \Biggl[ \Biggl({\dot a \over a} \Biggr)^2
+ {\ddot a \over a} \Biggr] = - \hat{\kappa}^2\,\hat{T}_{55}$.
One can see that accelerated expansion $\ddot a >0$, which
is required for inflation, takes place with a negative $\hat{T}_{55}$
such as that found in Eqns.(\ref{emgamma}) and (\ref{eq:pbulk}). 

In addition to the bulk Einstein equations, we have boundary
conditions (known as the Israel jump conditions)
due to the fact that our observable brane is embedded in the bulk.
We will assume that
the energy momentum tensor on the boundary can be written in a
perfect fluid form, 
$t^0_0 = \rho_{br}$
and 
$t^1_1= - p_{br}$,
where $\rho_{br}$ and $p_{br}$ are the energy density and
pressure, respectively, measured by a comoving observer.
For the metric in eq.(\ref{simple}) and 
the brane at a $Z_2$ symmetry fixed plane, 
the Israel conditions become
\ba
3[a'/a]_0 &=& -\hat{\kappa}^2 b_0 \rho_{br} \nn\\
3[n'/n]_0 &=& \hat{\kappa}^2 b_0 (2\rho_{br}+3 p_{br}) \, .
\ea
Here, subscript $0$ refers to quantities evaluated at the location
of the brane at $y=0$, and $[X] \equiv X{y\rightarrow 0^+} - X(y \rightarrow 0^-)$
indicates the "jump" in the quantity $X$ as one crosses the boundary (brane)
at $y=0$ (the value of $X$ as one approaches the brane from one side
minus the value of $X$ as one approaches it from the other side).

\section{Results: Behaviour of scale factors $a(t)$ and $b(t)$:
Inflation without Inflatons}

Our results are obtained by solving 
Einstein's equations
together with the constraints provided by the Israel conditions. 
We use the energy momentum tensor derived above in Eq.(\ref{eq:imp})
 appropriate to a primordial Hagedorn epoch.
First we will consider the case where
the brane is energy/pressureless and that $\dot{b}=0$
(where $b$ is the scale factor of the extra dimension).
We then catalogue more general types of behaviour that
can occur for $\dot{b}\neq 0$.

\subsection{Case I: 
$\Lambda_{br} = \Lambda_{bulk} =\dot{b}=0$}

Let us first impose
$\dot b=0$ simply as an external condition; \ie
the extra dimension is fixed in time.  
We also set the cosmological constants  both on the 
brane and in the bulk to zero, 
$\Lambda_{br} = \Lambda_{bulk}= 0$. All other 
contributions to the energy momentum localized on the brane 
(for example massless Yang-Mills degrees of freedom)
have an entropy that is subdominant 
to the limiting bulk degrees of freedom as long as the string
energy density 
$\rho> \rho_c$ (where $\rho_c$ is a critical Hagedorn density of order 1
in string units). 
For the moment we will therefore neglect the brane energy-momentum
and set $a_0'(t)=n_0'(t)=0$.
In this discussion the brane at $y=0$ 
is playing no role in determining the evolution of the
cosmology; the scale factor $a_0$ changes purely as a result of the 
bulk equations. 
It is simple to solve the 55 equation
for the scale factor $a_0(t)$ on our brane (at $y=0$).

\noindent\underline{ {\it Standard case with $d_o=0$:} }
In the high energy case with windings
in all transverse dimensions we 
see from Eqs.(\ref{eq:imp}) and (\ref{eq:pbulk})
that $T_5^5 \sim \sqrt{\rho} \sim a^{-3/2}$
(where, again, $\rho$ is the local energy density measured with 
respect to the volume $\Vp$)
and hence, 
\begin{equation}
\label{61}
a_0(t) \sim t^{4/3} \, ,
\end{equation}  
\ie {\em power law inflation}.
For the more general case of `p' large parallel dimensions
in any number of extra perpendicular dimensions, we find
$T_5^5 \sim a^{-p/2}$ and $a \sim t^{4/p}$.  
Here adiabaticity has been assumed in taking $E \sim S \sim {\rm const}$.
We note the amusing
result that inflation requires $p \leq 3$ (for our brane
$p=3$) and speculate on its role in our
three large extra dimensions.

\noindent\underline{ {\it Case with $d_o=2$:} }
In the case of 2 dimensions with no windings,
we find a period of  
{\em exponential inflation}:
\be 
\label{expinf}
\frac{a_0(t)}{a_0(0)} = \exp \left( 
{
-\frac{t^2\hat{\kappa}^2}
{8\Vt^{\frac{3}{2}}} +t \sqrt{\frac{\hat{\kappa}^2}{6 
\Vt^{\frac{3}{2} }} \log (\rho(0)e^{3/4})}  }\right) 
\ee
where $\rho(0)$ is the initial density at $t=0$. 
Initially the second term in the exponent dominates
and there is exponential inflation.
This solution has an automatic end to inflation 
when $\rho(t) \sim 1$ (in string units), just 
as the system is dropping out of the Hagedorn phase. 

\noindent\underline{ {\it  Other systems:}} 
We summarize the energy momentum tensors and cosmological 
behavior of $a_0(t)$ for systems with $d_o$
perpendicular dimensions without windings
in Table 1, where $p=3$ for our universe as a 3-brane. 
Superluminal expansion is found for systems with $d_o \leq 2$,
i.e., no more than two perpendicular dimensions without windings.
In all cases, we have also found
solutions to the $T_{00}$ and $T_{ii}$ equations.

In adiabatic systems the Hagedorn regime and hence the inflationary 
behaviour we have found eventually come to an end.  
For a system to be in the Hagedorn regime 
requires an entropy density higher than the 
critical Hagedorn density (of order 1 in string units).
Below this density the energy momentum tensor and hence 
the cosmology is governed by the massless 
relativistic Yang-Mills gas (the gas is
present on the brane even in the Hagedorn regime but is subdominant). 
Thus there is no problem exiting 
from the inflationary behaviour.  In fact the
main issue is how long inflation can last. We return to 
this question later when we discuss how inflation can be 
sustained.

\vspace{0.5cm}
\begin{table}[ht]
{\footnotesize 
\centerline{
\begin{tabular}{|c||c|c|c|}
\hline 
regime & 
$\rho(\beta)=E/\Vp$ & 
$- p_{bulk}$ &
$a_0(t)/a_0(0)$  \\
\hline
\hline
&&& \\
$d_o=0$ &
$\frac{1}{\Vt}(\beta-\beta_H)^{-2}$ & 
$\rho^{\frac{1}{2}}$ & 
$t^{\frac{4}{p}}$\\
$d_o=1$ &
$(\beta-\beta_H)^{-\frac{3}{2}}$ & 
$\rho^{\frac{1}{3}}$ & 
$t^{\frac{6}{p}}$ \\
$d_o=2$ &
$(\beta-\beta_H)^{-1}$ & 
$\log \rho$ & 
$\exp \left( {-C t^2 + D t }\right) $ \\
$d_0=3$ &
$ (\beta-\beta_H)^{-\frac{1}{2}} $ & 
$ \rho^{-1} $ & 
$ t^{-\frac{2}{p}} $ \\
$d_0=4$ &
$-\log (\beta-\beta_H)$ & 
$e^{- \rho}$ & 
const \\
&&& \\ \hline 
\end{tabular}
}}
\caption{Cosmological regimes for open strings in the Hagedorn phase
with $p$ large parallel dimensions, where $p=3$ for
our observable universe as a 3-brane. 
Here $d_0$ is the number of dimensions transverse to the brane
in which there are no windings; in our standard ``high-energy''
regime we take $d_0 = 0$.  Here $\rho$ is the energy density
of strings, \ie , the $T_{00}$ component of the bulk energy-momentum
tensor. In addition, $T_{55} = p_{bulk}$ is the negative bulk pressure
that drives inflation of our scale factor $a_0(t)$. Note
inflationary expansion 
for $d_o \leq 2$, i.e., at most 2 directions without windings.
The constants 
$C$ and $D$ are given for $p=3$ in eq.(\ref{expinf}).  }
\label{table1}
\end{table}

\subsection{Case II:  
$\frac{\hat{\kappa}^2}
{12}\Lambda_{br}^2 - \Lambda_{bulk} =0$ and $\dot{b}\neq 0$}

We also studied the case where the extra dimension is not stabilized 
but there is still no nett cosmological constant. 
By `nett' we mean that the contribution from $a'$ and 
$n'$ in $G_5^5$ 
cancels the bulk contribution on the RHS of the $T_5^5$ equation.
The condition for this is 
\be 
\Lambda_{nett} = -\frac{\hat{\kappa}^2}{12}\Lambda_{br}^2 + \Lambda_{bulk} =0.
\ee
Using the 55 equation, we catalogued
some possible types of behavior for $a$ with different ans\"atze
for $b$.
Since $\alpha\approx 1 $, for $\gamma=-1,-\half,\half$ ($\gamma = d_o/2-1$), 
the pressure is 
\begin{equation}
\label{eq:t55}
-p_{bulk} 
\propto a^{{-3\gamma}\over {\gamma-1}} b^{-3\over2} \, .
\end{equation}

Taking $\Lambda_{nett}=0$, we find a family of power law 
solutions for $\gamma=-1,-\half,\half$ of the form 
\ba
\label{eq:plaw}
a_0(t) &\approx & At^q\nn\\
b_0(t) &\approx & Bt^r 
\ea 
where subscript-0 indicates values at $y=0$, 
\be
q={ {{\gamma-1}\over {2\gamma}} ({4\over 3}-r)} ,
\ee
and $r$ is arbitrary.
For our standard $\gamma = -1$,
we find superluminal solutions with $q>1$ for $r<1/3$.
Hence one can have an inflating brane with a shrinking
bulk ($r<0$) or with a growing bulk ($0<r<1/3$).
We also find a family of hyperbolic solutions with
\be
{a_o(t) \over a_o(0)} = \Biggl({{\rm sinh}2C(t+t_1) \over
{\rm sinh}2Ct_1} \Biggr)^{1/2}.
\ee
These solutions can give an
exponentially increasing scale factor on our brane.
The proper choice of solution in Case II depends on the
initial value of $\dot b_o$.

\subsection{Case III:
$\frac{\hat{\kappa}^2}
{12}\Lambda_{br}^2 - \Lambda_{bulk} \neq 0$ and $\dot{b}\neq 0$}

In \cite{us}
we also consider the case of additional cosmological constants
in the brane and bulk. The results can be found in our longer paper.
Not surprisingly we recover the usual cosmological constant driven
inflation when we set $\rho=0$.

\section{Sustaining inflation and solving the Horizon Problem}

In the previous sections we have found the onset of inflation
via superluminal growth of the scale factor.  We also know that
eventually, once the temperature drops out of the Hagedorn
regime, inflation ends.  This graceful exit from inflation
is an appealing feature of this scenario.  
However, we cannot calculate in between.
Once inflation begins, our calculations break down; our
interpretation of small changes in the metric corresponding
to small changes in volume no longer holds.  
At this point we can only speculate 
what happens in between, and suggest two possibilities. 

First, previous authors have argued that strings in a de Sitter 
background are unstable to fluctuations and 
that this instability can sustain a period of de Sitter 
inflation~\cite{englert,turok,veneziano}. This phenomenon is well known 
for strings once they are in the de Sitter-like 
phase\footnote{We qualify `de Sitter' because the exponential 
solutions we have do not possess the full O(1,4) de Sitter symmetry.}. 
However the missing ingredient that the present study adds 
is an explanation for how the universe enters a de Sitter like phase 
in the first place.   Our mechanism gets the universe into
the locally de Sitter phase, whereupon the mechanism of
\cite{englert,turok,veneziano} keeps it there 
{\em without} having to assume adiabaticity.

Second, our 3-brane sits in a bath of branes and bulk strings
with ongoing dynamics such as branes smashing into each other
and possibly annihilating.  The dynamics is likely to keep
the system hot.  Hence high temperatures near the Hagedorn
temperature may be sustained for a long period; inflation
continues during this period.  We note that the high
entropy of our brane $S\sim 10^{88}$ may originate from
the dynamics of this hot brane/string bath.   Eventually
the system cools, drops out of the Hagedorn regime, and inflation ends.

There is a slight difference between our study and 
the results of \cite{turok} which we should comment on. 
In \cite{turok}, the density never goes {\em above}
a critical density, $\rho'_c$.
In our case, on the other hand, we must have $\rho > \rho_c \sim 1$ for the 
calculation to be valid (\ie to be in the Hagedorn regime).
The difference arises
because \cite{turok} was concerned with 
{\em stretched} strings and 
introduced a cut-off in the momentum integral.
This `coarse graining' put an artificial
upper limit on the amount of string that can be packed into the 
volume and effectively removed the Hagedorn behaviour.
Conversely, the present paper begins in the regime $\rho > \rho_c$ 
where the calculations of \cite{turok} end.


\section{Conclusion and discussion}

We have studied the possible cosmological implications of
the Hagedorn regime of open strings on D-branes in the weak coupling limit. 
Our main result is that a gas of open strings can exhibit negative pressure 
leading naturally to a period of power law or exponential inflation.
Hagedorn inflation also has a natural exit 
since any significant cooling can cause a change in the thermodynamics
if winding modes become  quenched or if the density drops 
below the critical density, $\rho_c\sim 1$, needed for the entropy 
of the Hagedorn phase to be dominant. 
To summarize, open strings on branes in the hot early Hagedorn phase
of the universe provide a mechanism to drive a period of
inflation, even in the absence of any potential. 

\subsection*{Acknowledgements}

\noindent We thank D. Chung, C. Deffayet, E.
Dudas, K. Olive, G. Servant and C. Savoy for discussions. 
S.A.A. and I.I.K. thank 
J. Barb\'on and E. Rabinovici for a previous collaboration
and discussions concerning this work.
S.A.A. thanks the C.E.A. Saclay for support during this work.
K.F. acknowledges support from the Department of
Energy through a grant to the University of Michigan.  K.F. thanks
CERN in Geneva, Switzerland and the Max Planck Institut fuer Physik in
Munich, Germany for hospitality during her stay. 
I.I.K. was  supported in part by PPARC rolling grant
PPA/G/O/1998/00567, the EC TMR grant FMRX-CT-96-0090 and  by the INTAS
grant RFBR - 950567. 



\begin{thebibliography}{99}
\frenchspacing
\def\prpts#1#2#3{Phys. Reports {\bf #1}, #2 (#3)}
\def\prl#1#2#3{Phys. Rev. Lett. {\bf #1}, #2 (#3)}
\def\prd#1#2#3{Phys. Rev. D {\bf #1}, #2 (#3)}
\def\prc#1#2#3{Phys. Rev. C {\bf #1}, #2 (#3)}
\def\plb#1#2#3{Phys. Lett. {\bf #1B}, #2 (#3)}
\def\npb#1#2#3{Nucl. Phys. {\bf B#1}, #2 (#3)}
\def\apj#1#2#3{Astrophys. J. {\bf #1}, #2 (#3)}
\def\apjl#1#2#3{Astrophys. J. Lett. {\bf #1}, #2 (#3)}

\bibitem{guth}
A. Guth, \prd{23}{347}{1981}

\bibitem{us}
S.A. Abel, K. Freese, and I.I Kogan,
{\it JHEP} {\bf 0101} (2001) 039

\bibitem{carlitz}
R. Hagedorn, {\it Suppl.~Nuovo Cimento} {\bf 3} (1965) 147;
S. Frautschi, \prd{3}{2821}{1971}; 
R.D. Carlitz, \prd{5}{3231}{1972}

\bibitem{general}
K. Huang and S. Weinberg,
\prl{25}{895}{1970};
E. Alvarez, \prd{31}{418}{1985}; \npb{269}{596}{1986}; 
M. Bowick and L.C.R. Wijewardhana, \prl{54}{2485}{1985};
B. Sundborg, \npb{254}{883}{1985};
S.N. Tye, \plb{158}{388}{1985};
P. Salomonson and B. Skagerstam, \npb{268}{349}{1986}; {\it Physica}
{\bf A158} (1989) 499; 
E. Alvarez and M.A.R. Osorio, \prd{36}{1175}{1987}; 
D. Mitchell and N. Turok, \prl{58}{1577}{1987}; \npb{294}{1138}{1987};  
I. Antoniadis, J. Ellis and D.V. Nanopoulos, \plb{199}{402}{1987}; 
M. Axenides, S.D. Ellis and C. Kounnas, \prd{37}{2964;}{1988}; 
I.I. Kogan, JETP. Lett. {\bf 45} (1987) 709;
B.Sathiapalan,  \prd{35}{3277}{1987};  
J. Attick and E. Witten, \npb{310}{291}{1988};
A.A. Abrikosov Jr. and Ya. I. Kogan
\ijmpa{6}{1501}{1991} (submitted 1989),
{\it Sov. Phys. JETP} {\bf 69} (1989) 235; 
R. Brandenberger and C. Vafa, \npb{316}{391}{1989};
M.J. Bowick and S.B. Giddings, \npb{325}{631}{1989};
S.B. Giddings, \plb{226}{55.}{1989};
B.A. Campbell, N. Kaloper, K.A. Olive, 
\plb{277}{265}{1992};
S.A. Abel, \npb{372}{189}{1992};
N. Kaloper, R. Madden, K.A. Olive, \npb{452}{677,}{1995};
M.L. Meana, M.A.R. Osorio and  J.P. Penalba,  
\plb{400}{275,}{1997},  
\plb{408}{183}{1997}.

\bibitem{deo}
N. Deo, S. Jain and C.-I. Tan, \plb{220}{125}{1989};
\prd{40}{2646}{1989}; N. Deo, S. Jain, O. Narayan and C.-I. Tan,
\prd{45}{3641}{1992}

\bibitem{thorl}
D.A. Lowe and L. Thorlacius, \prd{51}{665}{1995}, 
S. Lee and L. Thorlacius, \plb{413}{303}{1997}.

\bibitem{abkr}
J.L.F. Barb\'on, I.I. Kogan and E. Rabinovici, \npb{544}{1999}{104}, 
S.A. Abel, J.L.F. Barb\'on, I.I. Kogan, and E. Rabinovici,
JHEP {\bf 04} (1999) 015

\bibitem{after}
M.L. Meana and J.P. Penalba,
\npb{560}{154}{1999}; \plb{447}{59}{1999};
M.A. Vazquez-Mozo, \prd{60}{106010}{1999};
B. Sundborg, \hepth{9908001};
S.A. Abel, J.L.F. Barb\'on, I.I. Kogan, and E. Rabinovici,
\hepth{9911004}

\bibitem{polch}
For reviews see: J. Polchinski, {\em TASI} lecturs on D-branes, 
\hepth{9611050}; ``String Theory'', vols 1,2 (CUP) 1998.

\bibitem{KKOP} P. Kanti, I. I. Kogan, K. A. Olive, M. Pospelov.

\bibitem{moreref} 
K. Benakli, Int. J. Mod. Phys. {\bf D8} (1999) 153,
A. Lukas, B.A. Ovrut and D. Waldram, \hepth{9902071};
H.A. Chamblin and H.S. Reall, \npb{562}{133}{1999};
P. Bin\'etruy, C. Deffayet and D. Langlois, Phys.Lett. {\bf B544} (2002) 183;

\bibitem{chuf}
D. Chung and K. Freese, \prd{61}{023511}{2000}.

\bibitem{englert}
R. Brout, F. Englert and E. Gunzig, Ann. Phys. {\bf 115} (1978) 78;
Gen. Rel. Grav. {\bf 10} (1979) 1; R. Brout, F. Englert and P. Spindel, 
\prl{43}{417}{1979}; Y. Aharanov and A. Casher, \plb{166}{289}{1986};
Y. Aharonov, F. Englert and J. Orloff, \plb{199}{366}{1987}

\bibitem{turok}
N.Turok, \prl{60}{549}{1988}

\bibitem{veneziano}
N.~Sanchez and G.~Veneziano,
Nucl.\ Phys.\  {\bf B333} (1990) 253;
M.~Gasperini, N.~Sanchez and G.~Veneziano,
Int.\ J.\ Mod.\ Phys.\  {\bf A6} (1991) 3853;
Nucl.\ Phys.\  {\bf B364} (1991) 365;

\bibitem{riotto}
A. Riotto,
Phys.Rev. {\bf D61} (2000) 123506.

\end{thebibliography}
\end{document}